\documentclass[twocolumn,aps,prl]{revtex4}
\input{epsf}
\newcommand{\be}{\begin{eqnarray}}
\newcommand{\ee}{\end{eqnarray}}
\begin{document}
\title{\vspace*{0.5in}Possible extensions of the standard
cosmological model: anisotropy, rotation, and magnetic field
}

\author{M. Demia\'nski}
\address{Institute of Theoretical Physics, University of Warsaw,
00-681 Warsaw, Poland\\
        Department of Astronomy, Williams College,
           Williamstown, MA 01267, USA}
\author{A.G. Doroshkevich}
\address{Astro Space Center of Lebedev Physical
           Institute of  Russian Academy of Sciences,
                        117997 Moscow,  Russia\\}

\begin{abstract}
We show that the difference between the theoretically expected and
measured by WMAP amplitude of the quadrupole fluctuations of CMB can
be related to the impact of the anisotropic curvature of the
homogeneous universe dominated by the dark energy. In such universe
the matter expansion becomes practically isotropic just after the
period of inflation and only at small redshifts the anisotropic
expansion is generated again by the small curvature $\Omega_K=1-
\Omega_m- \Omega_\Lambda \leq 10^{-4}$. For such models the
possible deviations from the parameters derived for the standard
cosmological model are evidently negligible but the correlations of
large scale perturbations and distortions of their Gaussianity are
possible. Such models are also compatible with existence of a
homogeneous magnetic field and matter rotation which contribute to
the low $\ell$ anisotropy and can be considered as ``hidden
parameters'' of the model. Their influence can be observed as, for
example, special correlations of small scale fluctuations and
the Faraday rotation of the CMB and radiation of the farthest
quasars. However, both the magnetic field and matter rotation
require also modifications of the simple models of isotropic
inflation and they change the evolutionary history of the early
Universe.

PACS number(s): 98.80.Es.
\end{abstract}

\maketitle

\section{Introduction}
Recent rapid progress of observational cosmology resulted in
formulation of the ``Cosmological Standard Model'' (CSM) (see,
e.g., review by Springel, Frenk\,\&\,White 2006). This model
successfully describes main observations at redshifts $0\leq z
\leq 10^9$ such as the primordial nucleosynthesis, the CMB
fluctuations (WMAP, Hinshaw et al. 2006), the Large Scale
Structure in the observed galaxy distribution (SDSS, Tegmark
et al. 2004; 2dFGRS, Cole et al. 2005) and the redshift
variations of brightness of SN Ia (Astier et al. 2006;
Wood-Vasey et al. 2007).

The CSM provides us with a reasonable framework for its further
extensions which will change and improve interpretation of some
observational data but the well established characteristics of the
basic model will remain unchanged. Some of such extensions were
already considered during last years. Thus, Durrer, Kahniashvily\,
\&\,Yates (1998), Kosovsky et al. (2005), Durrer (2007) and 
Kahniashvily \& Ratra (2007) have considered the CMB anisotropy 
generated by the magnetic field and, in particular, by Alfven waves. 
A possible observational manifestation of small scale entropy 
perturbations were discussed in Naselsky and Novikov (2002) and 
Doroshkevich et al. (2003). Both extensions can weakly distort the 
primordial nucleosynthesis and fluctuations of the CMB temperature 
and polarization but they do not change the basic parameters of 
the CSM.

On the contrary, attempts of Jaffe et al. (2005, 2006), Cayon et al.
(2006), Bridges et al. (2006), McEwen et al. (2006)  to exploit the
Anisotropic Cosmological Models (ACM) with a large curvature to
improve the interpretation of many anomalies of the observed maps of
CMB fluctuations are evidently in contradiction with  both the
inflationary model and the observational upper limit for the spatial
curvature of the Universe in CSM.  More detailed analysis of this
model (Naselsky \& Verkhodanov 2006) reveals also serious
divergences between expected and observed phases of reconstructed
harmonics what further decreases the appeal of such attempts to
drastically change the interpretation of measured fluctuations.

The most active discussion is now concentrated around anomalies in
the large scale fluctuations of the CMB temperature measured by
WMAP. As is well known, for low $\ell$ multipoles there are serious
deviations from Gaussianity which can be mostly caused by
imperfections of the methods used for discrimination between the CMB
and foreground (see, e.g., Naselsky et al. 2005 and references
there). In particular,  the ``standard model'' predicts the
amplitude of quadrupole fluctuations at the level $\Delta T^2=1250
\mu K^2$ (Spergel et al. 2006) while the observed value provided by
both COBE and WMAP measurements is only $\Delta T^2=249 \mu K^2$
(Hinshaw et al. 2006). Moreover, in several papers it was declared
that the quadrupole and octopole moments are correlated with their
two preferred planes surprisingly closely aligned. The preferred
axis, dubbed the ``axis of evil'' by Land and Magueijo (2005), is
found to be pointing towards Virgo and is close to the ecliptic pole
(Oliveira-Costa et al. 2004). Detailed discussion of these problems
can be found in Oliveira-Costa \& Tegmark (2006) where more than
hundred references are summarized.

Here we cannot discuss these results and all the proposed so far
interpretations of these peculiarities. Let us only note that
the measured quadrupole fluctuations of the CMB are formed by a
combined action of the suitable {\it random} inhomogeneities in
the matter motion and distribution and of the possible {\it
regular} factors such as the anisotropic curvature, homogeneous
magnetic field and/or matter rotation. However, the extension of
the CSM by incorporating these {\it regular} factors stimulates
a renewed interest in the Anisotropic Cosmological Models.

The Anisotropic Cosmological Models (ACM) were quite popular in the
60$^{th}$ and 70$^{th}$ (see, e.g. Zel'dovich \& \,Novikov 1983)
when scarcity of observational information have not restricted such
discussions. As was established in many papers, for such models the
strong anisotropic expansion at an early stage is incompatible with
observational estimates of abundances of light elements. Early
anisotropic expansion can be naturally suppressed at the period of
inflation and after inflation the ACM becomes identical to the
isotropic Friedman-Robertson-Walker (FRW) model with a negligible
magnetic field and rotation. This means that the model proposed by
Campanelli et al. (2006) is incompatible with the standard models of
inflation (see also discussion below in Sec. 4).

However, the anisotropic curvature present in all ACM (with the
exception of Bianchi type I models) decreases but does not disappear
during the period of inflation. Almost all ACM with a suitably small
anisotropic curvature are close to the isotropic FRW model at
redshifts $z\gg 1$. In fact, the influence of the small
anisotropic curvature on the cosmological recombination of hydrogen
or the evolution of perturbations and galaxy formation is
negligible. So, it does not change all conclusions of CSM related to
the evolution of the Universe at $z\geq 1$ such as, in particular,
the small scale CMB fluctuations. However, at redshifts $z\leq 1$
the anisotropic curvature generates again the {\it regular} weak
anisotropy of the matter expansion which provides the {\it regular}
contribution to the CMB quadrupole.

Here we illustrate this statement by considering the simplest ACM
with the axial symmetry which provides the natural explanation of
the deficit of power in the quadrupole mode. Even this model
allows to introduce also the  homogeneous magnetic field as a
``hidden parameter'' of the cosmological model. More complex ACMs
allow also to introduce  rotation of the Universe generalizing the
well known cosmological model of G\"odel (1952) (see, e.g., Ellis
\,\&\, MacCallum 1969). Some published estimates related to such
more complex ACM are shortly summarized in Sec. 4.

\section{Anisotropic Cosmological Model}

The simplest ACM is the Bianchi type III model which was
discussed already in Schucking and Heckmann (1958), Kompaneets
and Chernov (1965) and Doroshkevich (1965). Of course, this
extension is not unique and more complex Bianchi models can be
also considered leading to similar results (see, e.g.,
Doroshkevich et al. 1972 a, b).

This ACM is described by the metric
\be
ds^2=c^2dt^2-a^2(t)[dr^2+f^2(r)d\phi^2]-b^2(t)dz^2\,,
\label{metric}
\ee
\be
f(r)=\{{\rm sh}\,r;\,\,r;\,\,\sin\,r\}\,,
\ee
for models with negative, zero and positive curvatures. Here
$c$ is the speed of light and the functions $a(t)\,\&\,b(t)$
describe the anisotropic expansion of the Universe.

For the standard matter the energy--momentum tensor is \be
T_{ik}=(\varepsilon+p)u_iu_k-pg_{ik}\,, \label{t_ik} \ee where
$\varepsilon\,\&\,p$ are the energy density and pressure of the
matter. In this case the evolution is determined by three
equations
\[
\dot{\alpha} + \alpha(2\alpha+\beta)+\delta a^{-2} = \Lambda +
(\kappa/2)(\varepsilon-p)\,,
\]
\be
\dot{\beta} + \beta(2\alpha+\beta) = \Lambda +
(\kappa/2)(\varepsilon-p)\,,
\label{eq}
\ee
\[
\alpha^2 + 2\alpha\beta+\delta a^{-2} = \Lambda +
\kappa\varepsilon\,,
\]
where dot $\dot{}$ denotes the time derivative and
\[
\alpha=\dot{a}/a,\quad \beta=\dot{b}/b,\quad \kappa=8\pi G/c^4\,.
\]
Here $\Lambda>0$ is the cosmological constant and $\delta=-1, 0,
1$ for the models with negative, zero and positive curvature
(\ref{metric}).

For the most interesting cases, namely, the anisotropic inflation in
the early Universe and matter dominated expansion ($p=0$) describing
the evolution of the Universe at redshifts $z\leq z_{eq} \approx
10^4$ simple solutions can be obtained.

\subsection{Period of the anisotropic inflation}

Thus, near the singularity, for $t\rightarrow 0$, we can neglect
the impact of the curvature and matter density and consider the
influence of the cosmological constant only. In this case we get
for $\Lambda=\Lambda_{inf}$ (Ellis\,\&\,MacCallum 1969;
MacCallum 1971)
\be
a(t)=a_0{\rm ch}^{2/3}\tau\cdot{\rm th}^{p_1}\tau,\quad
b(t)=b_0{\rm ch}^{2/3}\tau\cdot{\rm th}^{p_2}\tau\,,
\label{inflation}
\ee
\[
2p_1+p_2=2p_1^2+p_2^2=1,\quad \tau=t\sqrt{3\Lambda_{inf}}/2\,.
\]
This expression generalizes the well known Kasner solution by
including the impact of the cosmological constant $\Lambda_{inf}$.
At $\tau\leq 1$ these expressions describe the usual anisotropic
expansion which can be linked with the influence of the large
scale gravitational waves (see, e.g., Belinsky et al. 1972, 1982;
Doroshkevich e al. 1972 a, b;
Zel'dovich\,\&\,Novikov 1983; Landau\, \&\,Lifschitz 1991). This
anisotropic expansion is rapidly suppressed and at $\tau\geq 1$
we get almost isotropic expansion with
\[
a(t)\propto b(t)\propto \exp(2\tau/3),\quad a/b\rightarrow a_0/
b_0\,.
\]
Non the less, in this model the anisotropy of the spatial
curvature is retained. Moreover, the anisotropic expansion at
$\tau\leq 1$ can generate anisotropy of the large scale
perturbations what destroys their Gaussianity and produces
correlations caused by the existence of preferred directions in
the Universe.

Of course, these directions do not overlap with the dipole of
the CMB, or with the direction to the Virgo or to the ecliptic
pole and, so, these peculiarities cannot explain all the
correlations among the measured low $\ell$ multipoles of the
CMB. However, they demonstrate that the standard assumptions of
isotropy and Gaussianity of {\it all} primordial perturbations
could be too restrictive (see, e.g., Mukhanov 2005, Sec. 5).

\subsection{Period of the matter dominated expansion}

For the later period of the cosmological expansion when we can
neglect the impact of the pressure ($p\ll\varepsilon$) the
solutions of the system (\ref{eq}) can be written as follows:
\be
3\alpha^2=H_0^2[\Omega_m(1+z)^3+\Omega_\Lambda+\Omega_K(1+z)^2]\,,
\label{alpha}
\ee
where $H_0\,\&\,z=a(0)/a(t)-1$ are the Hubble
constant and the redshift measured in the plane of symmetry,
$\Omega_m\,\&\, \Omega_\Lambda$ are dimensionless density of the
Dark Matter and Dark Energy and
\[
\Omega_K=1-\Omega_m-\Omega_\Lambda,\quad \Omega_K\propto -\delta\,,
\]
characterizes the curvature of the Universe. Integration of
(\ref{alpha}) links the scale factor $a$ with time $t$.
Evidently, at the matter dominated stage, $(1+z)^3\geq
\Omega_\Lambda/\Omega_m$, we have $a\propto t^{2/3}$ and we
recover the isotropic solution.

The anisotropy of expansion is measured by the ratio $a/b$: \be
(\alpha-\beta)\dot{}+(\alpha-\beta)(2\alpha+\beta)=-\delta/a^2\,.
\label{zaxis}
\ee
For small $|\Omega_K|\ll 1$ and $|a/b-1|\ll 1$ we
get \be \ln\left({a\over b}\right)\approx\Omega_K\int_{1+z}^\infty{
\xi^2d\xi\over\sqrt{\Omega_m\xi^3+\Omega_\Lambda}} \int_\xi^\infty
{\eta^{-2}d\eta\over\sqrt{\Omega_m\eta^3+ \Omega_\Lambda}}\,.
\label{psi} \ee For $(1+z)^3\gg (\Omega_\Lambda/\Omega_m)\geq 0$ we
have
\[
\ln(a/b)=0.4(\Omega_K/\Omega_m)(1+z)^{-1}\,,
\]
what corresponds to the growing mode of the density perturbations.
For the accepted in CSM values of
\be
\Omega_\Lambda\approx 0.7,\quad \Omega_m\approx 0.3,\quad
(\Omega_m/\Omega_\Lambda)^{1/3}\approx 0.754\,,
\label{omega}
\ee
we get from (\ref{psi}) for $z=0$ and $T\approx 2.7 K$
\[
\Theta_K=\ln(a/b)\approx a/b-1\approx 0.77\Omega_K,\quad
|\Omega_K|\ll 1\,,
\]
\be \Delta T/T=\{-\Theta_K/3,\, -\Theta_K/3,\, 2\Theta_K/3\}\,,
\label{dt/t} \ee for the relative temperature fluctuations along the
coordinate axes (\ref{metric}). As it follows from
(\ref{zaxis},\,\ref{dt/t}) we have $\Theta_K =0$ for the flat model
with $f(r)=r,\, \delta=\Omega_K=0$.

This result shows that the small positive or negative curvature
provides us with the reasonable quadrupole anisotropy of the CMB
temperature. In galactic coordinates the orientation of this
quadrupole is arbitrary while its amplitude depends upon $\Omega_K$.
The required curvature is quite small because the expected
$|\Theta_K|\sim|\Omega_K|\sim\sqrt{\langle (\Delta
T/T)^2\rangle}\sim 10^{-5}$. For negative curvature $\Omega_K>0$,
the expansion along $z$-axis is slower than that in the plane of
symmetry and $\Theta_K>0$. For small positive curvature
$\Omega_K<\,0$, the expansion along $z$-axis is faster than that in
the plane of symmetry and $\Theta_K<\,0$.

At the same time it is evident that the required curvature
$|\Omega_K|<\,10^{-5}$ is within range of $\sim$10\% precision
achieved by available measurements of cosmological parameters (see,
e.g., Hinshaw et al. 2006). 

\section{Quadrupole anisotropy of CMB}

As is well known, the five quadrupole coefficients are linked with
the components of a symmetric traceless tensor $Q_{ij}$ \be
{Q_{11}\over\sqrt{2}}=a_{2,2}-{a_{2,0}\over\sqrt{6}},\,\,
{Q_{12}\over\sqrt{2}}=-a_{2,-2},\,\, {Q_{13}\over\sqrt{2}}
=-a_{2,-1}\,, \label{quadr} \ee
\[
{Q_{22}\over\sqrt{2}}=-a_{2,2}-{a_{2,0}\over\sqrt{6}},\quad
{Q_{23}\over\sqrt{2}}=-a_{2 ,1},\quad {Q_{33}\over\sqrt{2}}
=2\,{a_{2,0}\over\sqrt{6}}\,.
\]
This representation allows us to find the principle values of this
tensor, $\lambda_i$, and their orientation in the galactic
coordinates $l\,\&\,b$. With the standard definitions we have for
the 3 years ILC measurements (Hinshaw et al. 2006):
\[
\lambda_1=~~27.1 \mu K,\quad (l,b)=(-0.8^\circ\pm 13^\circ,\,\,
63.3^\circ\pm 1^\circ)\,,
\]
\be \lambda_2=~~12.9 \mu K,\,\, (l,b)=(15.5^\circ\pm 3^\circ,\,\,\,
25.8^\circ\pm 1.2^\circ), \label{ilc3} \ee
\[
\lambda_3=-40~~\mu K,\quad (l,b)=(-77.6^\circ\pm 5^\circ,\,\,\,
6.5^\circ\pm 4^\circ)\,,
\]
with
\[
I_2/5=-(\lambda_1\lambda_2+\lambda_1\lambda_3+\lambda_2\lambda_3)/5
=\Delta T^2=250 \mu K^2\,.
\]
These orientations differ from both the dipole direction
\[
(l,b)_D=(-96^\circ,48^\circ)\,,
\]
and the possible but quite arbitrary orientation of the quadrupole
introduced by Oliveira-Costa et al. (2004) \be
(l,b)=(-110^\circ,60^\circ)\,. \label{toh2} \ee The actual
orientations of the observed octopole and higher multipoles can be
reconstructed also using the principle axes of corresponding
symmetric tensors.

The observed quadrupole is a superposition of the regular and the
random ones. The regular quadrupole is related to the anisotropic
expansion of the Universe while the random one is generated by the
initial perturbations. Relative orientation of these quadrupoles is
arbitrary and, so, we can obtain only the rough estimate of the
required regular quadrupole and the curvature of the Universe.
Comparing the observed and expected quadrupoles we get \be
|\Theta_K|\approx |\Omega_K|\sim 10^{-5}- 10^{-4}\,. \label{res} \ee
For example, for the expected value $\Delta T^2=1250\mu K$,
$\lambda_i$ as given by (\ref{ilc3}) and for the two special
orientations of the regular quadrupole (\ref{metric}) along the
principle axes (\ref{ilc3}) we get instead of (\ref{dt/t}) for the
relative quadrupole fluctuations:
\[
{\Delta T\over T}=\left\{{\lambda_1\over\sqrt{5}T}+{2\Theta_1
\over 3},{\lambda_2\over\sqrt{5}T}-{\Theta_1\over 3},
{\lambda_3\over\sqrt{5}T}-{\Theta_1\over 3}\right\}\,,
\]
\be \Theta_1=\Theta_K\approx\{14.6;\,\,\,-28.1\}\cdot 10^{-6},
\label{Thet1} \ee
\[
{\Delta T\over T}=\left\{{\lambda_1\over\sqrt{5}T}-{\Theta_3
\over 3},{\lambda_2\over\sqrt{5}T}-{\Theta_3\over 3},
{\lambda_3\over\sqrt{5}T}+{2\Theta_3\over 3}\right\}\,,
\]
\be \Theta_3=\Theta_K\approx\{32.5;\,\,\,-12.7\}\cdot 10^{-6}\,.
\label{Thet3} \ee This choice of the quadrupole orientation is
consistent with the possible correlations of the low $\ell$
perturbations and the spatial curvature discussed in Sec. 2.1\,.
 These results are consistent with the available estimates of
the cosmological parameters (see, e.g. Spergel et al. 2006) and
weakly depend upon the accepted values of $\Omega_\Lambda$ and
$\Omega_m$.

\section{Models with the magnetic field and rotation}

The small anisotropic curvature can only generate the large
scale anisotropy of the CMB. However, combined with the matter
rotation and/or the magnetic field such anisotropy will also
successfully distort the Gaussianity of small scale CMB fluctuations
and, in particular, generate correlations between different modes of
these fluctuations. Already available restrictions of the admissible
amplitude of the rotation and magnetic field show that, as a rule,
their interactions with the cosmological plasma leads to several
observable effects. However, potentialities of such interactions are
not exhausted and, in spite of their incompatibility with the
standard inflation models, they deserve further discussion.

\subsection{Influence of the magnetic field}

As was shown in Zel'dovich (1965), Doroshkevich (1965), and Thorne
(1967) the model (\ref{metric}) can include a homogeneous magnetic
field along the $z$ axis. Presently accepted estimates of this field
$B(z=0)\leq 10^{-9}G$ are obtained from the observed regular
magnetic fields within galaxies ($\sim 10\mu G$) and clusters of
galaxy ($\sim 0.1 \mu G$) (see reviews by Widrow 2002; and Semikoz
\& Sokoloff 2005). However, such fields do not noticeably influence
the expansion rate of the Universe and the CMB fluctuations because
their energy density, $w(z)=B^2(z)/8\pi\propto (1+z)^4$, is small in
comparison with the energy density of CMB, $\varepsilon_{rad}
\approx 4\cdot 10^{-13}(1+z)^4erg/cm^3$, \be \Omega_B=\kappa
c^2{w(0)\over 3H_0^2}\leq 10^{-11}\left({B(0)\over
10^{-9}G}\right)^2,\,\, {w\over\varepsilon_{rad}}\approx 10^5
\Omega_B\,. \label{H} \ee

Indeed, the contribution of the weak magnetic field is described by
the term $\Omega_B(1+z)^4$ in Eq. (\ref{alpha}) and instead of
(\ref{psi}) we will have for the function $\Theta=\Theta_B$ \be
\Theta_B=2\Omega_B\int_{1+z}^{1+z_r}{\xi^2d\xi\over\sqrt{
\Omega_m\xi^3+\Omega_\Lambda}}\int_\xi^{1+z_r}{d\eta\over
\sqrt{\Omega_m\eta^3+\Omega_\Lambda}}, \label{theta-h} \ee where
$z_r\approx 10^3$ is redshift of the hydrogen recombination and for
$z\geq z_r$ the radiation can be considered as isotropic. As is seen
from (\ref{theta-h}), the amplitude of regular quadrupole is \be
\Theta_B\approx {4\Omega_B\over 3\Omega_m}z_r\approx 4\Omega_B\cdot
10^3\leq 4\cdot 10^{-8}\left({B(0)\over 10^{-9} G}\right)^2\,,
\label{TB} \ee and at least a field $B\geq 10^{-7}G$ is required to
reproduce the expected regular quadrupole (\ref{Thet1},\ref{Thet3})\,.
More details can be found in Durrer (2007) and  Kahniashvili \& Ratra 
(2007).

\subsection{Influence of the matter rotation}

Many homogeneous models are compatible also with the matter rotation
(see, e.g., Ellis\,\&\,MacCallum 1969; Barrow et al. 1985). The
evolution of slow rotation is quite similar to that of the magnetic
field and is determined by the angular momentum conservation law
(see, e.g., Landau\,\&\,Lifschitz 1991). Thus, for the matter
dominated Universe at $z\leq z_{eq}\approx 10^4$ the angular
velocity of matter, $\omega(z)$, is changing as \be {\omega(z)\over
H(z)}\approx\sqrt{1+z\over\Omega_m}\,\, {\omega(0) \over H_0}\,,
\label{mrot} \ee where $\omega^2(z)=\omega_\nu(z)\omega^\nu(z)$ and
$\omega_\nu$ and $\omega^\nu$ are the covariant and contravariant
components of the angular velocity. This relation shows that at the
matter dominated period the slow rotation does not influence the
evolution of the Universe. At the radiation dominated period for
$z\geq z_{eq}\approx 10^4$ we have \be \omega(z)/H(z)\propto
(1+z)^{-1}\,, \label{rot} \ee and, so, the maximal value of
$\omega/H$ is achieved at $z\sim z_{eq}\approx 10^4$. However, the
rotation influences the evolution of the models near the singularity
(see, e.g., Ellis\,\&\,MacCallum 1969).

The observed rotation of galaxies corresponds to
\[
\omega(0)/H_0\leq 10^{-2}\,,
\]
but the measured CMB anisotropy leads to stronger restrictions.
Thus, for one of the Bianchi VI models discussed in Ellis \&
MacCallum (1969) we get for the quadrupole temperature fluctuations
\be \Theta_\omega=\Theta_K-{2z_r^3\over
9\Omega_m}\left({\omega(0)\over H_0}\right)^2\,, \label{theta_rot}
\ee where $\Theta_K\,>0$ is given by (\ref{dt/t}) and $z_r\approx
10^3$ is the redshift of recombination epoch. It is interesting that
in this case the influence of the curvature and rotation can partly
compensate each other and, therefore, $\Omega_K\,>10^{-5}$ and
$\omega(0)/H_0\,>10^{-7}$ can be considered.

More complex pattern appears for the Bianchi VII model where
together with quadrupole fluctuations also higher multipoles
correlated with the quadrupole are generated. Detailed analysis of
these fluctuations (Barrow et al. 1985; Jaffe et al. 2005, 2006;
Ghosh et al. 2006) leads to the following estimates
\[
\omega(0)/H_0\leq 4\cdot 10^{-5} - 10^{-6}\,,
\]
and even in this model we can expect that $\omega(z)/H(z)\ll 1$ at
all redshifts.

These results indicate that rotation and/or magnetic field can
influence the evolution of the Universe mainly during the
radiation dominated period at redshifts $z\geq z_{eq}\approx
10^4$ or at the period of cosmological recombination $z\sim
z_r\approx 10^3$.

\subsection{Observational manifestations}

The most interesting observational manifestations of both factors
are the possible rotation of the CMB polarization, distortions of
its Gaussianity and the power spectra of the temperature and
polarization at small and large scales, and, in particular,
intermixture of the ``E'' and ``B'' modes of the CMB polarization
(see, e.g., Kosowsky\,\&\,Loeb 1996; Harari et al. 1997; Campanelli
et al. 2004; Kosowsky et al. 2005; Subramanian 2006; Giovannini
2006). In fact, the expected Faraday rotation of the CMB is
estimated to be
\be
\langle\varphi^2\rangle^{1/2}\approx 1.^\circ6\left(B\over 10^{-9}G
\right)\left(\lambda\over 1cm\right)^2\,.
\label{frot}
\ee
Another observational manifestation of the magnetic field is the
Faraday rotation of the radiation of farthest quasars (see, e.g.,
Widrow 2002; Hutsemekers et al. 2005; Semikoz \& Sokoloff 2005).

Durrer et al. (1998) have considered the distortions of small
scale CMB fluctuations caused by the interaction of CMB with the
possible Alfven waves and even find similar correlations in the
observed temperature maps of WMAP (Durrer 2007; Kahniashvili \&
Ratra 2007). However, such identifications are in question 
(Naselsky et al. 2004).  

It is well known that both the rotation and magnetic fields are
strongly suppressed in the models of standard inflation (see, e.g.,
Widrow 2002). This means that serious discussions of ACM with
rotation and magnetic field require suitable modifications of the
inflationary models, or alternatively generation of the rotation
and/or magnetic field in the course of evolution during or after
inflation (see, e.g., Turner and Widrow 1988; Ratra 1992;
Lasenby\,\&\,Doran 2004).

\section{Discussion}

The ``Cosmological Standard Model'' with inflation at the
earlier period of cosmological expansion and domination of the
cosmological constant at small redshifts significantly extends
the class of cosmological models compatible with presently
available observational restrictions. In particular,  the recent
discussion of the high precision WMAP measurements (see, e.g.,
Oliveira--Costa \& Tegmark 2006 and references there) points out
that, perhaps, some of the observed peculiarities can be
successfully interpreted in the framework of anisotropic
cosmological models rather than in the Friedman-Robertson-Walker
cosmology. It is also important that such homogeneous
anisotropic models allow to include into consideration the
primordial rotation and magnetic fields.

The anisotropic cosmological models were popular in the 60$^{th}$
and earlier 70$^{th}$ when attention was mostly focused on models
that incorporated the Friedman-Robertson-Walker universe with the
isotropic expansion (Zel'dovich \,\&\,Novikov 1983). Now we see
(\ref{inflation}) that the simple models of inflation successfully
suppress the initial anisotropy linked with an expansion, magnetic
field or rotation and provide almost isotropic expansion of the
Universe down to small redshifts.

On the other hand, at small redshifts the cosmological constant
efficiently suppresses the impact of the moderate anisotropic
curvature and makes the expansion almost isotropic for a wide
class of Bianchi models. Our estimates (\ref{psi},\,\ref{dt/t},
\,\ref{res} -- \ref{Thet1}) demonstrate that the required
deviations from the flat isotropic Universe could be quite small
and their observational manifestations invite further careful
study. Non the less, the consequences of such extensions are
quite important because now the wide set of Bianchi models can
be again used to explain various observed manifestations of the
large scale anisotropy. In particular, in such models we can
expect that
\begin{enumerate}
\item{} Distortions of Gaussianity of large scale perturbations
and the anisotropy of their power spectrum lead to corresponding
distortions of low $\ell$ random multipoles and their possible
correlations.
\item{} For models with small spatial curvature the regular
quadrupole term dominates.
\item{} For the VII$_0$ Bianchi model with rotating matter
the regular octopole and higher multipoles correlated with
quadrupole are generated.
\item{} The magnetic field generates noticeable non
Gaussianity in the small scale CMB fluctuations and leads to
intermixture of the ``E'' and ``B'' modes of CMB polarization.
\end{enumerate}

In the flat Bianchi type I model the anisotropy is related to
the initial anisotropic expansion only and it rapidly
increases with redshift. So, for this model only negligible
anisotropy is compatible with the standard inflationary models
and the cosmological nucleosynthesis. Because of this restriction
and as was found above (Sec. 4.1) the simple elliptical model
discussed in Campanelli et al. (2006) cannot provide the adequate
explanation of the anisotropy of the CMB. At the same time,
discussion in Naselsky\,\&\,Verkhodanov (2006) demonstrates
the limited potentialities of the anisotropic cosmological
models with the large spatial curvature (Jaffe et al. 2005,
2006; Cayon et al. 2006; Bridges et al. 2006; McEwen et al.
2006) for description of the WMAP observations of low $\ell$
multipoles.

Up to the present time analysis of the CMB fluctuations
provides us with the best limitations of possible distortions of the
standard cosmological model. In spite of such limitations the
discussed extension of the cosmological models allows us to include
into consideration as ``hidden'' parameters not only the
anisotropic curvature and the entropy perturbations but, perhaps,
also the magnetic field and rotation of the Universe, what implies
various observational manifestations discussed above (Sec. 4). These
effects could be revealed by high precision measurements of the
Planck mission and detailed high precision observations of distant
quasars. However, we still do not have  reasonable models of
inflation that allow generation of magnetic fields  (see review in
Widrow 2002; Semikoz\,\&\,Sokoloff 2005) and the rotation during or
just after inflation.

\section{Acknowledgments}
This paper was supported  in part by the Polish State Committee
for Scientific Research grant Nr.1-P03D-014-26 and Russian Found
of Fundamental Investigations grant Nr.05-02-16302. AD thanks
Yu. Gnedin, B. Komberg, V. Lukash and P. Naselsky for useful
comments.
We should like to thank the anonymous referee for useful comments.

{}

\end{document}